\documentclass[%
superscriptaddress,
preprint,
amsmath,amssymb,
aps,
pra,
]{revtex4-1}

\usepackage{epsfig}

\usepackage{graphicx}
\usepackage{subfigure}
\usepackage{dcolumn}
\usepackage{bm}
\usepackage{longtable}
\usepackage{threeparttablex}
\usepackage{amssymb}
\usepackage{amsthm}
\usepackage{amsmath}
\usepackage{lineno}
\usepackage{hyperref}
\usepackage{adjustbox}
\usepackage{wasysym}
\usepackage{multirow}
\usepackage{longtable}

\begin{document}
	
	
	\title{Polarization-dependent laser resonance ionization of beryllium}
		
	\author{Ruohong Li} 
	\email{ruohong@triumf.ca}
	\affiliation{TRIUMF-Canada's Particle Accelerator Centre, Vancouver, BC, Canada, V6T 2A3}              
	\author{Maryam Mostamand}
	\affiliation{TRIUMF-Canada's Particle Accelerator Centre, Vancouver, BC, Canada, V6T 2A3}
	\affiliation{Department of Physics and Astronomy, University of Manitoba, Winnipeg MB, Canada, R3T 2N2}
	\author{Jekabs Romans} 
	\affiliation{TRIUMF-Canada's Particle Accelerator Centre, Vancouver, BC, Canada, V6T 2A3}
	\affiliation{University of Applied Sciences Emden/Leer, 26723 Emden, Germany}              
	\author{Jens Lassen}
	\affiliation{TRIUMF-Canada's Particle Accelerator Centre, Vancouver, BC, Canada, V6T 2A3}
	\affiliation{Department of Physics and Astronomy, University of Manitoba, Winnipeg MB, Canada, R3T 2N2}
	\affiliation{Department of Physics, Simon Fraser University, Burnaby, BC, Canada, V5A 1S6}

	
   \begin{abstract}
	
	Using TRIUMF’s off-line laser ion source test stand with a system of tunable titanium sapphire lasers, the polarization dependence of laser resonance ionization has been investigated using beryllium. A significant polarization dependence was observed for the excitation path $^1$S$_0$$\rightarrow$$^1$P$^{\circ}_1$$\rightarrow$$^1$S$_0$, which are typical transitions for alkaline and alkaline-like elements. This polarization dependence was further verified on Be radioactive isotopes at TRIUMF's isotope separator and accelerator facility (ISAC). Laser polarization was proven to be an important parameter in operating resonance ionization laser ion sources (RILIS).
	
	The polarization spectroscopy was preformed off-line both on the 2p$^2$ $^1$S$_0$ autoionizing (AI) state and high-$n$ Rydberg states of the $2sns$ $^1S_0$ and $2snd$ $^1D_2$ series. The energy of the 2p$^2$ $^1$S$_0$ AI state and ionization potential (IP) of beryllium were extracted as 76167(6)~cm$^{-1}$ and 75192.59(3)~cm$^{-1}$. Polarization spectroscopy can be used to determine the $J$ values of newly found states in in-source spectroscopy of the complex/radioactive alkaline-like elements such as Ra, Sm, Yb, Pu and No.
	
	
    \end{abstract}

	\maketitle
	
	\section{Introduction}\label{Introduction}
	
	In a multi-step photoexcitation process, the excitation efficiency from the ground state to a final state depends on the polarizations of excitation photons and the total angular momentum of the intermediate and final states. This is particularity significant for the alkaline and alkaline-like element transitions $^1$S$_0$$\rightarrow$$^1$P$^{\circ}_1$$\rightarrow$$^1$S$_0$. Due to the polarization relaxation by collisions with environmental atoms, the reported studies of this effect were typically performed on low-density media such as atomic beams \cite{Bekov81,Yu89,Santala,Locke,Seema,Nit,Kim,Yi,Miyabe}, instead of a confined and hot environment like a hot-cavity laser ion source. Initially motivated by the observation of polarization dependence of beryllium (Be) ion yields during online radioactive beam delivery at TRIUMF's isotope separator and accelerator facility (ISAC), a systematic investigation of this effect was performed at the off-line laser ion source test stand (LIS-STAND) \cite{Lav13,Li13} with atoms in a hot-cavity environment. 
	
	Laser ionized Be radioactive ion beams are in high demand at ISAC: $^7$Be proton capture is the key process to understand the solar neutrino deficit; $^{11}$Be is the simplest halo nucleus with only one halo neutron; the neutron-rich nucleus $^{14}$Be with two halo neutrons is a loosely bound state of three bodies, similar to $^{11}$Li but has its special characteristics. High ionization efficiency will help to extract $^{14}$Be beams in measurable quantities for experiments. In addition, It is interesting for trace analytical studies measuring the small abundance of long live radioisotope $^{11}$Be in geological samples. The preferred ionization scheme for beryllium using Ti:Sa lasers is $2s^2$ $^1$S$_0$$\rightarrow$ $2s2p$ $^1$P$^{\circ}_1$$\rightarrow$ $2p^2$ $^1$S$_0$, which strongly depends on the polarizations of the exciting laser light theoretically when polarization relaxation is absent. Investigations of this polarization effect in a hot-cavity environment, such as resonance ionization laser ion sources, will be essential to provide high ion yields of these radioactive beams to various highly demanded experiments.       
	
	This polarization effect varies regarding to the hyperfine interaction between electron angular momentum and nuclear spins, and can be applied to separate isotopes using designed laser ionization schemes. Bekov \textit{et al.} selectively ionized even and odd Yb isotopes by multi-step photoionization using polarized laser beams \cite{Bekov81}. Santala \textit{et al.} used laser polarization to enrich odd-isotope of Gd \cite{Santala}. Polarization-based isotope‑selective photoionization of palladium and samarium isotopes were preformed \cite{Locke, Seema}. This spectroscopic technique was also used to determine the $J$ values in atomic structure studies \cite{Nit,Kim, Yi, Miyabe}. All these reported studies were preformed with stable isotope samples and at low atom density - atomic beams. However if verified in hot-cavity laser ion source, this polarization technique can also be applied to online in-source spectroscopy to identify $J$ values for newly found levels of alkaline-like elements, especially the complex atoms with only radioactive isotopes.

	\section{Theory}\label{Theory}
	In a multi-step photoexcitation process, the excitation efficiency from the ground state to a final state depends on the polarization vectors of excitation photons and the total angular momentum $J$ ($F$ if considering nuclear spin) of the levels involved in the transitions. For a case of n-step of photoexciation, the transition probability is proportional to the product of squared modulus of the transition dipole matrix elements~[\citenum{Yu89}]:
	
	\begin{equation}\label{eq1}
	\begin{aligned}
	W&\propto\prod_{i=1}^n |\left\langle n_il_iJ_iM_i |D_\lambda|n_{i-1}l_{i-1}J_{i-1}M_{i-1}\right\rangle |^2  \\
	&=\prod_{i=1}^n \left| \left(  
	\begin{array}{ccc}
	J_i & 1 & J_{i-1} \\
	-M_i & \lambda_i & M_{i-1}
	\end{array}
	\right) \times\left\langle n_il_iJ_i||D||n_{i-1}l_{i-1}J_{i-1}\right\rangle\right|^2\
	\end{aligned}
	\end{equation}

	where i=1, 2, $\cdots$ represents 1st, 2nd, $\cdots$ step of the excitation. $\lambda_i$ is equal to 0 for linear polarization and $\pm$1 for right/left circular polarizations of the laser photons. To satisfy the conservation of the angular momentum in the atom and light system, the Wigner's $3j$ symbol ${\left(  
		\begin{array}{ccc}
		J_i & 1 & J_{i-1} \\
		-M_i & \lambda_i & M_{i-1}
		\end{array}\right)}$ does not vanish only when M$_i$=$\lambda$$_i$+M$_{i-1}$. Therefore $\lambda_i$ can also be considered as the change of magnetic quantum number $M$ by absorption of photon. 
	
	For the same laser ionization scheme of an element, the reduced matrix element of the transition dipole moment ${\left\langle n_il_iJ_i||D||n_{i-1}l_{i-1}J_{i-1}\right\rangle}$ is the same for different polarization combinations of laser beams. Therefore, the polarization dependence of the excitation probability is determined by the product of the $3j$ symbols of the n-steps. The total probability is obtained by summation over all the possible paths $\sum W$.

	\begin{figure}[h]
		\begin{center}
			\includegraphics[width=0.9\textwidth]{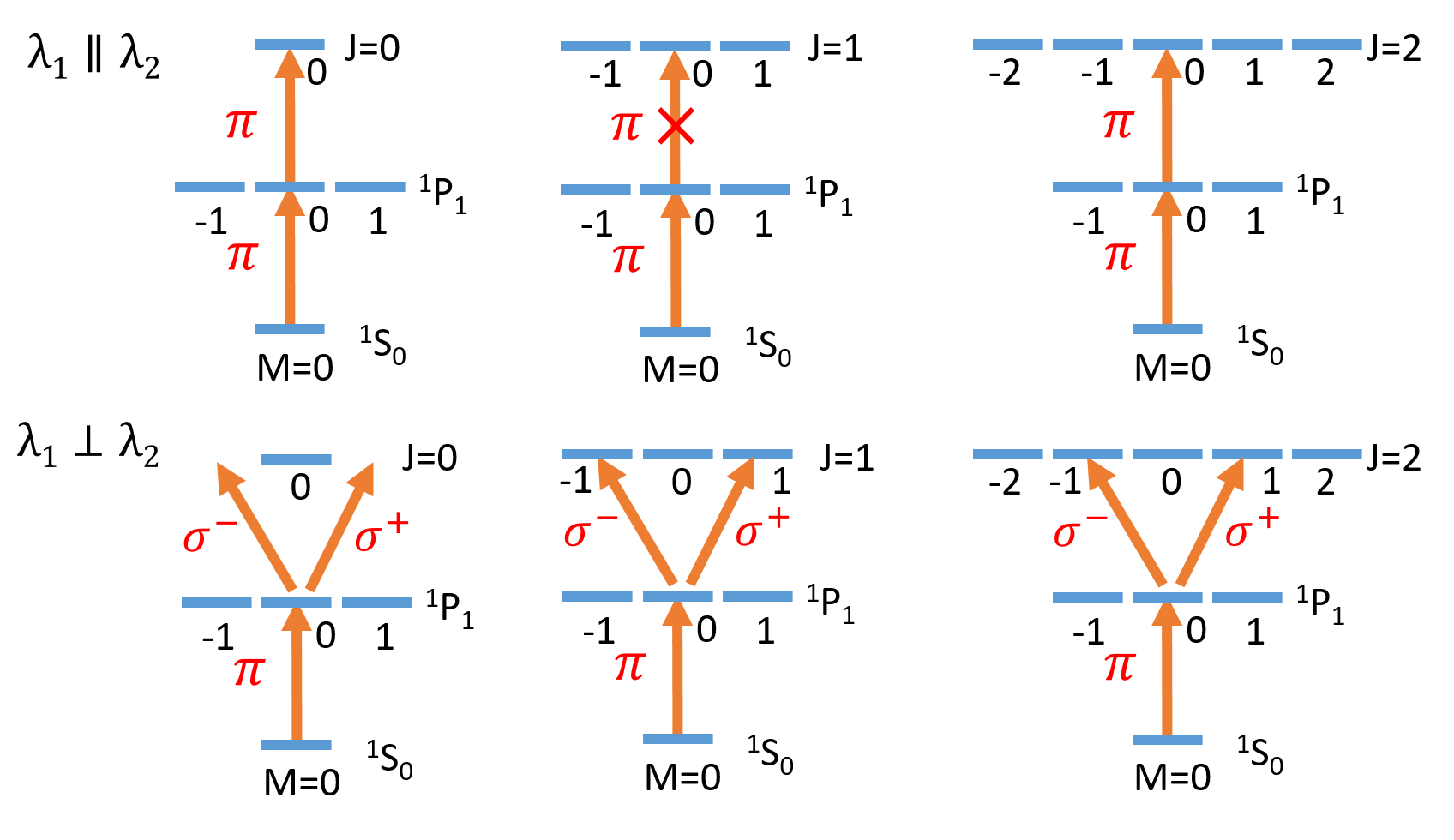}
		\end{center}
		\centering
		\caption{Polarization effect on the $6s^2$ $^1$S$_0$$\rightarrow$ $5d6p$ $^1$P$^{\circ}_1$$\rightarrow$ ${(J=0, 1,2)}$ scheme of barium. For the case of $\lambda_1$ $||$ $\lambda_2$, the excitation to the J=1 state is forbidden; for the case of $\lambda_1$ $\perp$ $\lambda_2$, the excitation to the J=0 state is forbidden.}
		\label{polarization}
	\end{figure}

	Considering the typical excitation path $6s^2$ $^1$S$_0$$\rightarrow$ $5d6p$ $^1$P$^{\circ}_1$$\rightarrow$ ${(J=0, 1,2)}$ of barium for example, only the magnetic sublevel $M=0$ can be populated if $\lambda_1$ is linearly polarized (Fig.~\ref{polarization}). If the polarization of $\lambda_1$ is parallel to that of $\lambda_2$, denoted as $\lambda_1$ $||$ $\lambda_2$, the final states with ${J=0, 2}$ are allowed. The $J=1$ state was forbidden due to the selection rule of $M_i=0$ $\nrightarrow$ $M_{i'}=0$ when $\bigtriangleup$$J=0$. In the case of $\lambda_1$ $\perp$ $\lambda_2$, $\lambda_2$ can be decomposed in to the superposition of right ($\sigma^+$) and left ($\sigma^-$) circularly polarized light when considering the $\lambda_1$ polarization direction as the quantization axis. As shown in Fig.~\ref{polarization}, the excitation from $^1P^{\circ}_1$ to a $J=0$ state is then forbidden, but those to ${J=1,2}$ states are allowed. This phenomenon has been well observed in Ba cross beam laser ionization spectroscopy~[\citenum{Baig09}]. The restriction shall be also valid for similar excitation paths in other alkaline and alkaline-like elements. 
	
	The above theory is valid only when the nuclear spin of the atom is zero, such as $^{138}$Ba. For isotopes with nuclear spin $I\neq0$, hyperfine interaction produces coupling between angular momentum $J$ and nuclear spin $I$, therefore $M_J$ will not be a good quantum number any more and Eq.1 should be rewritten as: 
	
	\begin{equation}\label{eq2}
	\begin{aligned}
	W\propto&\prod_{i=1}^n (2F_i+1)(2F_{i-1}+1)\left\lbrace 
	\begin{array}{ccc}
	J_{i-1} & F_{i-1} & I \\
	F_i & J_i & 1
	\end{array}
	\right\rbrace ^2\\
	&\left(  
	\begin{array}{ccc}
	F_i & 1 & F_{i-1} \\
	-M_i & \lambda_i & M_{i-1}
	\end{array}
	\right)^2\times|\left\langle n_il_iJ_i||D||n_{i-1}l_{i-1}J_{i-1}\right\rangle|^2 
	\end{aligned}
	\end{equation}
	
	Similar to Eq.1, the reduced matrix element is same for a given excitation scheme, therefore the polarization dependence can be calculated from Wigner's $3j$ and $6j$ symbols, which have a nuclear spin dependence. 
	
	This will be especially important for the laser ion source on-line operation to deliver radioactive isotope beams, since both even and odd isotopes have to be efficiently ionized. Bekov \textit{et al.} investigated the difference in the yields of even and odd Yb isotopes in multi-step photoionization by polarized laser beams \cite{Bekov81}. Even with a laser linewidth greater than the isotope shift, the polarization dependence of the odd and even isotope yields was evident.

	\section{Off-line experimental setup, procedure and results for Be}\label{Results}
	
\begin{figure}[ht]
		\begin{center}
			\includegraphics[width=0.6\textwidth]{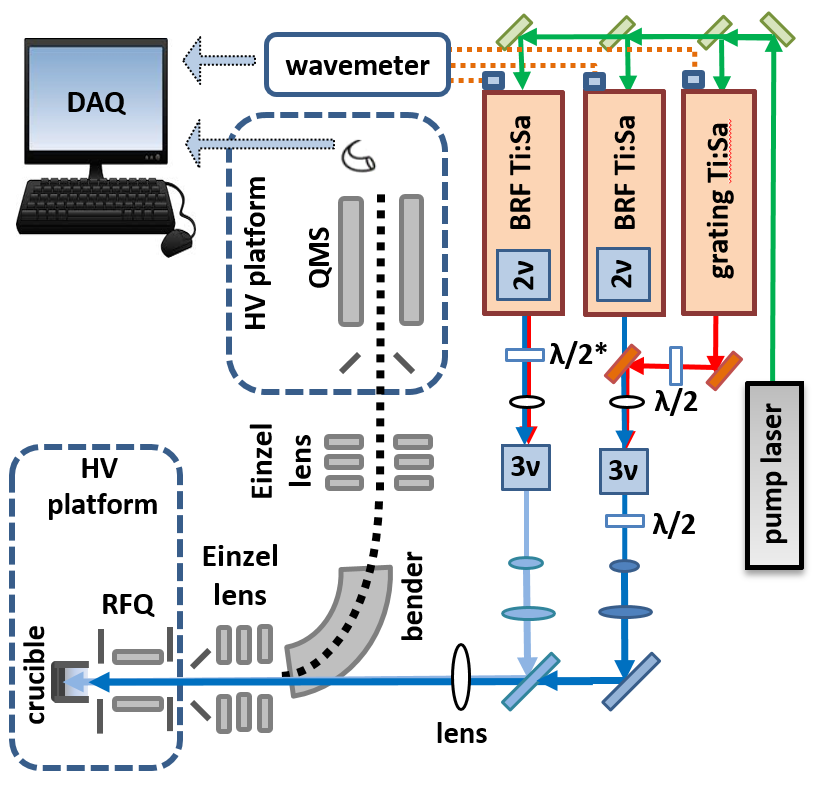}
		\end{center}
		\centering
		\caption{Schematic experimental setup for scheme A (see Fig.~\ref{scheme}) that consists of the laser system and the off-line test stand. RFQ: radio frequency quadrupole~[\citenum{Seb14}], QMS : quadrupole mass spectrometer, DAQ: data acquisition system, $\lambda$/2: half waveplate, $\lambda$/2*: dual-wavelength half waveplate, 2$\nu$: nonlinear crystal for frequency doubling, 3$\nu$: nonlinear crystal for frequency tripling.}
		\label{setup}
	\end{figure}

\begin{figure}[ht]
		\begin{center}
			\includegraphics[width=0.6\textwidth]{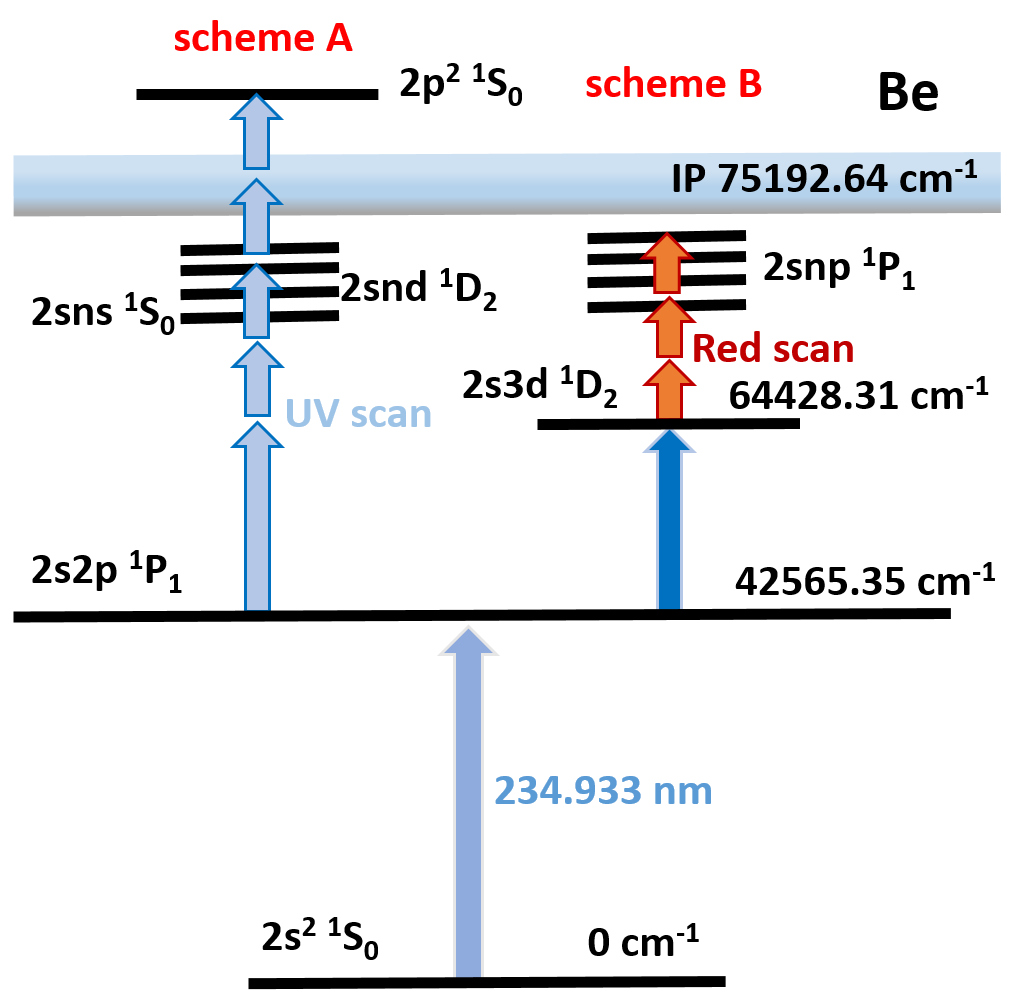}
		\end{center}
		\centering
		\caption{Laser resonance ionization schemes for beryllium investigated for polarization effects.}
		\label{scheme}
	\end{figure}

	The off-line experiment on stable $^9$Be was set up (Fig.~\ref{setup}) primarily for the two-step scheme, 2s$^2$ $^1$S$_0$ $\xrightarrow{\lambda_1=234.933~nm}$ 2s2p $^1$P$^{\circ}_1$ $\xrightarrow{\lambda_2=297.405~nm}$ 2p$^2$ $^1$S$_0$ (scheme A in Fig.~\ref{scheme}). The laser system consist of three Ti:Sa lasers simultaneously pumped by a 10~kHz 30~W Q-switched Nd:YAG laser (LEE laser LDP-100MQG). $\lambda_1$ was provided by frequency tripling of one birefringent-filter-tuned Ti:Sa laser. The fundamental (FD) and second harmonic (SH) frequencies were produced at the same time in the laser via an intra-cavity frequency doubling setup with the output coupler having 80$\%$ reflectivity at the FD wavelength and high transmission ($\sim$99$\%$) at the SH wavelength. The laser output powers were $\sim$1.4~W of FD and $\sim$ 300~mW of SH. In order to achieve type I frequency mixing inside a BBO crystal, a dual-wavelength waveplate was used to rotate SH light polarization by 90$^\circ$ while leaving the FD light polarization unchanged. An uncoated $f=$75~mm focusing lens was used to focus both FD and SH into the BBO crystal. This method removes the need for sensitive alignment to overlap the focal points of FD and SH inside the tripling crystal, and provided a concise setup of frequency tripling with moderate conversion efficiency. The output power of third harmonic (TH) light was 9~mW. 

	To reach the resonance wavelength of the 2nd step (297.405~nm) with fair laser power and obtain tunability at ultra violet (UV) range, two lasers were frequency mixed: a intra-cavity frequency doubled BRF Ti:Sa laser and a grating-tuned Ti:Sa laser \cite{Tei10}. To boost the output power of the grating laser, a 40$\%$ partial reflect mirror was applied inside the laser cavity \cite{Li_optics}. The output power of the UV light was 5-30~mW depending on the wavelength. A half waveplate was placed right after the frequency mixing system to allow polarization variation of the linearly polarized UV light.
	
	For studying polarization effects on the three-step scheme (scheme B in Fig.~\ref{scheme}), 2s$^2$ $^1$S$_0$ $\xrightarrow{\lambda_1=234.933~nm}$ 2s2p $^1$P$^{\circ}_1$ $\xrightarrow{\lambda_2=457.395~nm}$ 2s3d $^1$D$_2$ $\xrightarrow{IR~scan}$, the laser setup was slightly changed: $\lambda_2$ was generated directly from the intra-cavity doubling BRF laser, and the IR scan was provided by the FD light of the grating laser. 

	All lasers were expanded to 3-4$\times$ before they were superposed by 2-inch diameter dichroic mirrors. The combined laser beams were focused into the 3~mm diameter Ta crucible in the test stand chamber by a $f$= 5~m uncoated lens. The temporal superposition of the laser pulses was achieved by adjusting the pump power and using intra-cavity Pockel cells. The laser wavelengths were monitored by a High Finesse WS/6 wavemeter with a precision of 10$^{-6}$, which was routinely calibrated to a polarization stabilized HeNe laser with a wavelength accuracy of 10$^{-8}$ (Melles Griot 05 STP 901/903).  
	
	Alfa Aesar standard solution, 1~$\mu$g/$\mu$l Be in 5\% HNO$_3$ solution, was used as sample. About 50~$\mu$l solution was loaded into the Ta crucible, which could be resistively heated up to 2100~$^{\circ}$C inside the test stand vacuum system. The laser beams were focused and overlapped in the crucible and interacted with the atom vapor. The ions generated were accelerated to 10~keV, electrostatically bent and eventually focused into a quadrupole mass separator (QMS) to select the mass of interest. The filtered ion signal was detected by a channel electron multiplier at the end of the QMS. A detailed description of the test stand can be found in \cite{Lav13}.
	
	\subsection{Polarization dependence of resonance laser ionization via the 2p$^2$ $^1$S$_0$ AI state}
	
\begin{figure}[!htbp]
		\begin{center}
			\includegraphics[width=0.7\textwidth]{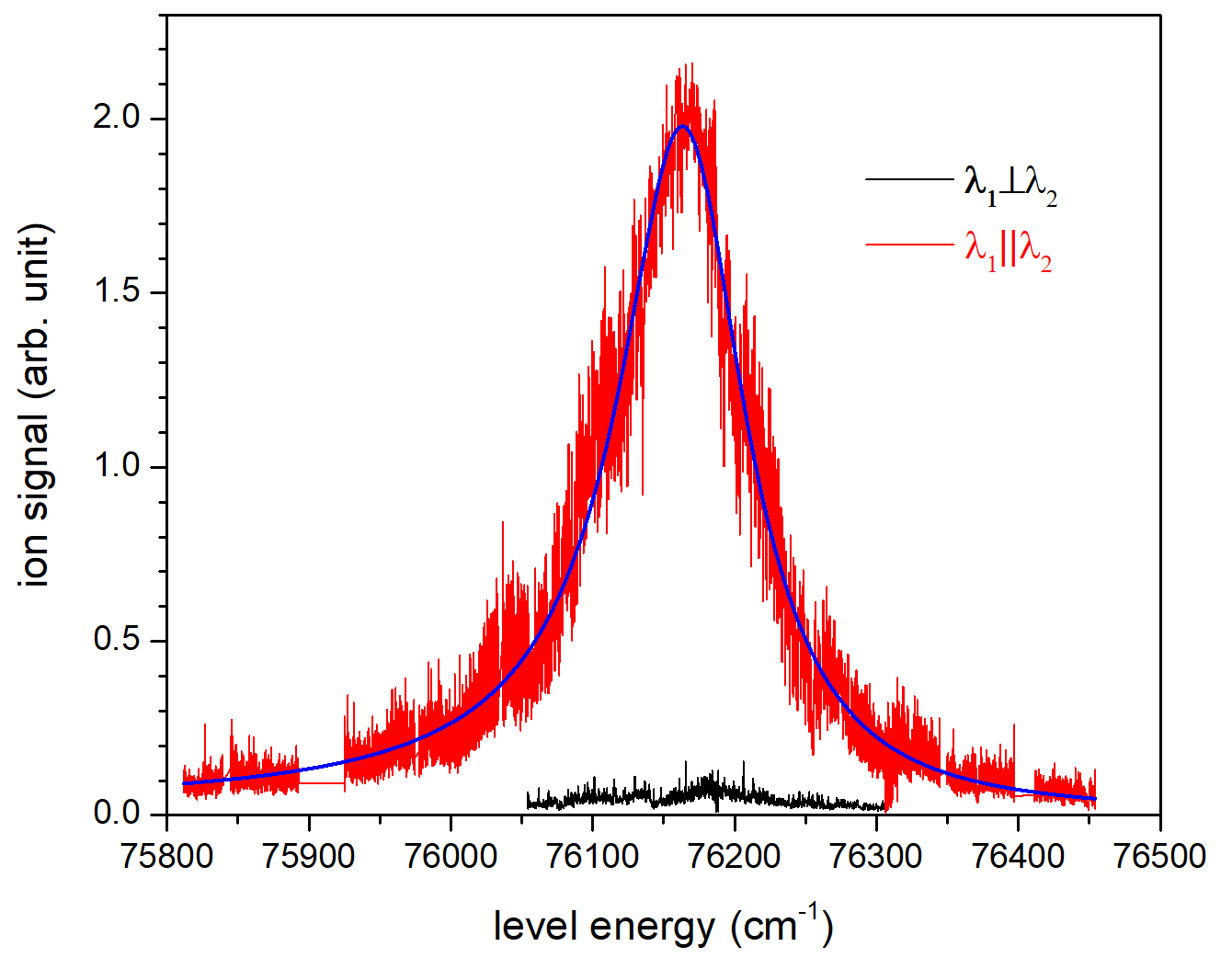}
		\end{center}
		\centering
		\caption{The 2p$^2$ $^1$S$_0$ AI state photoionization spectra of beryllium at different polarization orientations of the exciting lasers. The fitted center of the resonance is 76167(6)~cm$^{-1}$ using a Fano profile.} 
		\label{AI}
	\end{figure}

	The Be ion signal was first obtained and optimized using scheme A via the 2p$^2$ $^1$S$_0$ AI state. Photoionization spectra of the 2p$^2$ $^1$S$_0$ AI resonance using different polarizations of $\lambda_2$ were measured (Fig.~\ref{AI}). When $\lambda_1$ $||$ $\lambda_2$, the ion signal was high; in contrast when $\lambda_1$ $\perp$ $\lambda_2$, the ion signal almost vanished. The ratio of ion signals at the two polarization conditions is $\sim$22$\times$. This polarization dependence of ionization efficiency, although reported many times under low atom density conditions such as atomic beams \cite{Bekov81,Yu89,Santala,Locke,Nit,Seema,Kim,Yi,Miyabe}, has now also been observed in the hot-cavity environment of a laser ion source. Therefore laser polarization was proven to be an important operation parameter that deserves attentions for RILIS online beam delivery. This also implies that depolarization of the intermediate state is not significant in typical RILIS operations. Depolarization of excited atomic states could be caused by radiation diffusion or collisions with neighboring atoms \cite{Yakonov_1,Yakonov_2}. The radiation diffusion depolarization was theoretically predicted to become important at atomic densities greater than 10$^{10}$~cm$^{-3}$, and the collisional depolarization at higher concentration \cite{Yu89}. Some previous polarization dependence experiments using atomic beams specifically noted necessary quantitative control on the atom density at 10$^8$-10$^{10}$~cm$^{-3}$ to avoid depolarization \cite{Yu89,Nit}. 
	
	Limited by the volatility of the Be sample, the ionizer temperate was kept at 1500 $^\circ$C during the AI spectrum scans. However later during the online test at the ISAC-TRIUMF facility this polarization effect was also observed at the typical ISOL laser ion source temperature of $\sim$2000$^\circ$C. The details are presented in Sect.~\ref{online}.

	The observed 2p$^2$ $^1$S$_0$ resonance centers at 76167(6)~cm$^{-1}$ fitted by a Fano profile \cite{Fano65}, which is different from the NIST-quoted value of 76190(5)~cm$^{-1}$ \cite{NIST,Clark85}. The resonance width (FWHM) is $\sim$110~cm$^{-1}$. The spectral scan was taken when $\lambda_1$ power highly saturated the 1st step transition; but $\lambda_2$ power was not enough to saturate the AI transition. A saturation curve of $\lambda_2$ showed a linear relationship up to the maximum available UV laser power of 30~mW. Meanwhile the AI resonance is as wide as 400~cm$^{-1}$, that corresponds to 15~nm scan range for the grating laser, the laser power may vary and consequently affect the resonance profile. The variance of laser power was measured across the resonance. There was only $\pm$1~mW fluctuation on an average laser power of 29~mW. The resultant systematic error is not significant compared to the statistical fluctuation observed in the spectrum. To accommodate the obvious asymmetric figure of the AI resonance, a Fano profile was used in the fitting, which gave the resonance center at 76167(6)~cm$^{-1}$. To investigate the possible errors from  $\lambda_2$ frequency drifts and scan directions, three individual scans were made at different days. The results agreed with each other within 4~cm$^{-1}$.
	
	\begin{figure}[!htbp]
		\begin{center}
			\includegraphics[width=0.7\textwidth]{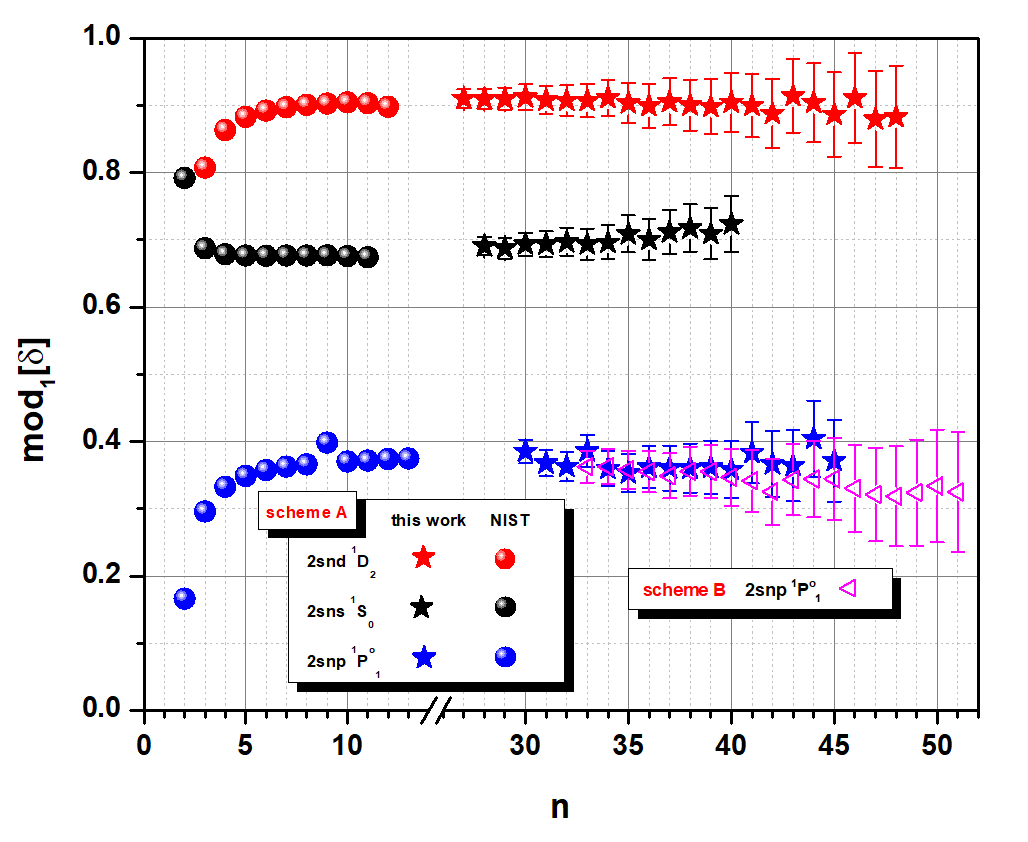}
		\end{center}
		\centering
		\caption{Mod$_1$[$\delta$] v.s. principal quantum number $n$ for three observed Rydberg series converging to the IP. Both the data of this work ($\bigstar$ for scheme A, and $\lhd$ for scheme B) and NIST data ($\newmoon$) are plotted. Different series $2snd$ $^1D_2$, $2sns$ $^1S_0$ and $2snp$ $^1P^{\circ}_1$ are colored red, green and blue respectively. The plotted $\delta$ were calculated using the new IP value of 75192.59(3)~cm$^{-1}$ obtained in this work, which is lower than currently accepted value 75192.64(6)~cm$^{-1}$ \cite{NIST,Beigang83}.} 
		\label{Be Lu}
	\end{figure}

	\subsection{Polarization dependent Rydberg spectra}
	
\begin{figure*}[!h]
		\begin{center}
			\includegraphics[width=1\textwidth]{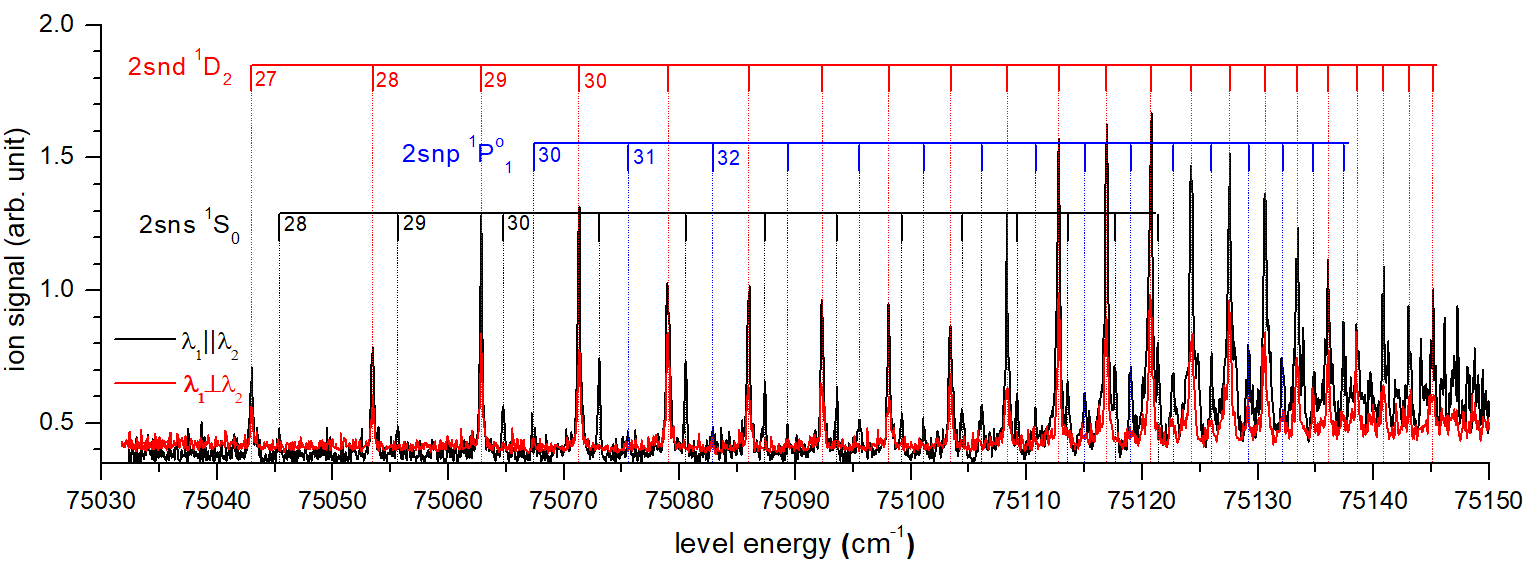}
		\end{center}
		\centering
		\caption{Rydberg spectra of beryllium using scheme A. Each Rydberg series is marked with a comb above the spectra. The $ns$ $^1S_0$ series vanishes when the polarizations of two exciting lasers are perpendicular to each other.} 
		\label{Rydberg}
	\end{figure*}

	Rydberg spectra were also obtained using different polarization orientation via scheme A (Fig.~\ref{Rydberg}). The comparison of the spectra at two conditions shows clearly the excitation path to $J=0$ was forbidden when $\lambda_1$ $\perp$ $\lambda_2$, same as the AI state of 2p$^2$ $^1$S$_0$. The observation of the $2snd$ $^1D_2$ resonance series with similar intensity under both parallel and perpendicular polarization orientation eliminated the speculations about possibly different transport efficiency of used optics at different polarizations.

	Excited from the $2s2p$ $^1P^{\circ}_1$ intermediate state, the Rydberg series $2sns$ $^1S_0$ and $2snd$ $^1D_2$ are allowed by the electric dipole selection rule. However additional forbidden Rydberg series $2snp$ $^1P^{\circ}_1$ also showed in the spectra: starting from $n$=30 upward for the spectrum of $\lambda_1$ $||$ $\lambda_2$; and from $n$=40 upward for the spectrum of $\lambda_1$ $\perp$ $\lambda_2$. The observation of these forbidden lines can result from Stark effects caused by the ambient stray electric fields from the ion extraction optics \cite{Bochchiat}. For high-$n$ Rydberg states, the mixing of different $l$ states by Stark effect becomes prominent even in electric fields as low as 100~ V/cm \cite{Kuhn}, which gives rise to violation of selection rules. To verify the $np$ $^1P^{\circ}_1$ series, a different laser excitation path (scheme B) was investigated. The observed members of the $np$ $^1P^{\circ}_1$ series (pink $\lhd$) are plotted in Fig.~\ref{Be Lu}. The IP values derived from these two schemes show agreement within experimental uncertainty. This also corroborates the accuracy and precision of the auto-scan UV spectra, since scheme B used the traditional infrared (IR) scan excited from a known intermediate state.
	
	\begin{table}[!h] \footnotesize
		\caption{Beryllium even-parity Rydberg series $2snd$ $^1D_2$ converging to the IP 75192.59(3)~cm$^{-1}$ and derived quantum defect $\delta$. The uncertainty in level energies is 0.15~cm$^{-1}$.}
		\begin{center}
			\begin{threeparttable}
				\setlength{\tabcolsep}{8pt}
				\renewcommand{\arraystretch}{0.8}
				\begin{tabular}{cccc}
					\hline\hline
					$n$	& \multicolumn{2}{c}{$\sigma$(cm$^{-1}$)}& $\delta$\\
					\hline	
					&this work&	NIST     &  \\
					\cline{2-3}
					3	&		&	64428.31	&	-0.19	\\
					4	&		&	68780.86	&	-0.14	\\
					5	&		&	71002.34	&	-0.12	\\
					6	&		&	72251.27	&	-0.11	\\
					7	&		&	73017.42	&	-0.10	\\
					8	&		&	73519.81	&	-0.10	\\
					9	&		&	73866.67	&	-0.10	\\
					10	&		&	74115.88	&	-0.10	\\
					11	&		&	74301.40	&	-0.10	\\
					12	&		&	74443.30	&	-0.10	\\
					$\cdotp$&$\cdotp$&$\cdotp$&$\cdotp$\\							
					27	&	75042.97	&		&	-0.08	\\
					28	&	75053.44	&		&	-0.08	\\
					29	&	75062.83	&		&	-0.08	\\
					30	&	75071.29	&		&	-0.08	\\
					31	&	75078.99	&		&	-0.08	\\
					32	&	75085.96	&		&	-0.08	\\
					33	&	75092.30	&		&	-0.08	\\
					34	&	75098.07	&		&	-0.07	\\
					35	&	75103.42	&		&	-0.08	\\
					36	&	75108.30	&		&	-0.08	\\
					37	&	75112.76	&		&	-0.07	\\
					38	&	75116.91	&		&	-0.08	\\
					39	&	75120.73	&		&	-0.08	\\
					40	&	75124.25	&		&	-0.07	\\
					41	&	75127.54	&		&	-0.07	\\
					42	&	75130.63	&		&	-0.08	\\
					43	&	75133.39	&		&	-0.05	\\
					44	&	75136.07	&		&	-0.06	\\
					45	&	75138.58	&		&	-0.08	\\
					46	&	75140.84	&		&	-0.05	\\
					47	&	75143.08	&		&	-0.08	\\
					48	&	75145.11	&		&	-0.07	\\								
					\hline\hline
				\end{tabular}
			\end{threeparttable}
		\end{center}
		\label{table Rydberg_nd}
	\end{table} 
	
	\begin{table}[!h] \footnotesize
		\caption{Beryllium even-parity Rydberg series $2sns$ $^1S_0$ converging to the IP 75192.59(3)~cm$^{-1}$ and derived quantum defect $\delta$. The data uncertainty is the same as Tab.~\ref{table Rydberg_nd}.}
		\begin{center}
			\begin{threeparttable}
				\setlength{\tabcolsep}{8pt}
				\renewcommand{\arraystretch}{0.8}
				\begin{tabular}{cccc}
					\hline\hline
					$n$	& \multicolumn{2}{c}{$\sigma$(cm$^{-1}$)}& $\delta$\\
					\hline	
					&this work&	NIST       &  \\
					\cline{2-3}
					2	&		&	0.00	&	0.79	\\
					3	&		&	54677.26	&	0.69	\\
					4	&		&	65245.33	&	0.68	\\
					5	&		&	69322.20	&	0.68	\\
					6	&		&	71321.15	&	0.68	\\
					7	&		&	72448.28	&	0.68	\\
					8	&		&	73146.57	&	0.68	\\
					9	&		&	73608.50	&	0.68	\\
					10	&		&	73930.40	&	0.68	\\
					11	&		&	74163.40	&	0.67	\\
					$\cdotp$&$\cdotp$&$\cdotp$&$\cdotp$\\							
					28	&	75045.37	&		&	0.70	\\
					29	&	75055.61	&		&	0.70	\\
					30	&	75064.74	&		&	0.70	\\
					31	&	75073.03	&		&	0.70	\\
					32	&	75080.52	&		&	0.71	\\
					33	&	75087.36	&		&	0.71	\\
					34	&	75093.57	&		&	0.71	\\
					35	&	75099.18	&		&	0.73	\\
					36	&	75104.44	&		&	0.72	\\
					37	&	75109.17	&		&	0.73	\\
					38	&	75113.56	&		&	0.74	\\
					39	&	75117.66	&		&	0.73	\\
					40	&	75121.37	&		&	0.75	\\					
					\hline\hline
				\end{tabular}
			\end{threeparttable}
		\end{center}
		\label{table Rydberg_ns}
	\end{table}
	
	\begin{table}[!h] \footnotesize
		\caption{Beryllium odd-parity Rydberg series $2snp$ $^1P^{\circ}_1$ converging to the IP 75192.59(3)~cm$^{-1}$ and derived quantum defect $\delta$. Due to the better signal to noise ratio, the level energies for $n$$\geqslant$33 were extracted from the spectrum via scheme B, and the level energies for $n$=30-32 were extracted from the spectra via scheme A. The data uncertainty is the same as Tab.~\ref{table Rydberg_nd}.}
		\begin{center}
			\begin{threeparttable}
				\setlength{\tabcolsep}{8pt}
				\renewcommand{\arraystretch}{0.7}
				\begin{tabular}{cccc}
					\hline\hline
					$n$	& \multicolumn{2}{c}{$\sigma$(cm$^{-1}$)}& $\delta$\\
					\hline	
					&this work&	NIST       &  \\
					\cline{2-3}
					2	&		&	42565.35	&	0.17	\\
					3	&		&	60187.34	&	0.30	\\
					4	&		&	67034.70	&	0.33	\\
					5	&		&	70120.49	&	0.35	\\
					6	&		&	71746.09	&	0.36	\\
					7	&		&	72701.80	&	0.36	\\
					8	&		&	73309.70	&	0.37	\\
					9	&		&	73709.40	&	0.40	\\
					10	&		&	74009.20	&	0.37	\\
					11	&		&	74221.10	&	0.37	\\
					12	&		&	74380.70	&	0.37	\\
					13	&		&	74504.10	&	0.38	\\
					$\cdotp$&$\cdotp$&$\cdotp$&$\cdotp$\\							
					30	&	75067.38	&		&	0.40	\\
					31	&	75075.56	&		&	0.38	\\
					32	&	75082.87	&		&	0.38	\\
					33	&	75089.49	&		&	0.38	\\
					34	&	75095.52	&		&	0.38	\\
					35	&	75101.06	&		&	0.38	\\
					36	&	75106.14	&		&	0.37	\\
					37	&	75110.81	&		&	0.37	\\
					38	&	75115.07	&		&	0.38	\\
					39	&	75119.02	&		&	0.38	\\
					40	&	75122.71	&		&	0.37	\\
					41	&	75126.12	&		&	0.37	\\
					42	&	75129.32	&		&	0.35	\\
					43	&	75132.19	&		&	0.38	\\
					44	&	75134.92	&		&	0.38	\\
					45	&	75137.47	&		&	0.38	\\
					46	&	75139.89	&		&	0.37	\\
					47	&	75142.14	&		&	0.36	\\
					48	&	75144.24	&		&	0.36	\\
					49	&	75146.19	&		&	0.37	\\
					50	&	75148.02	&		&	0.38	\\
					51	&	75149.77	&		&	0.38	\\					
					\hline\hline
				\end{tabular}
			\end{threeparttable}
		\end{center}
		\label{table Rydberg_np}
	\end{table}
	\subsection{IP determination}
	
	The currently accepted IP value in NIST database, 75192.64(6)~cm$^{-1}$, was obtained from similar step-wise laser resonance ionization spectroscopy by Beigang \textit{et al.} \cite{Beigang83} in 1983. Their wavelengths were calibrated via Rb and Cs cell and the level energies were claimed with accuracy of 3$\times$10$^{-6}$, which is $\sim$0.23~cm$^{-1}$ in the $n$$\geqslant$30 Rydberg region. Before Beigang's experiment, the most precise IP value was determined by Seaton as 75192.5(1)~cm$^{-1}$ \cite{Seaton}, based on Johansson's experimental data of $2sns$, $2snd$ and $2snf$ series \cite{Johansson}. Since only using low $n$ levels, Seaton's calculation included some theoretical corrections on quantum defects perturbed by $2pnl$ series. 
	
	The uncertainty of measured levels in this work is 0.15~cm$^{-1}$ which has been systemically verified by multiple experiments of different elements such as Sb, Lu and Po \cite{LiSb17,LiLu17,Seb18}. All measured Rydberg states are listed in Tab.~\ref{table Rydberg_nd}-\ref{table Rydberg_np}. The presented NIST data in the tables and Fig.~\ref{Be Lu} was measured by Johansson \cite{Johansson}. Using the Rydberg series $2snd$ $^1D_2$ observed via scheme A and one series $2snp$ $^1P^{\circ}_1$ observed via scheme B, the IP value of beryllium can be extracted precisely via Rydberg-Ritz formula: 
	\begin{equation}
	E_{n}=IP-\frac{R_{M}}{n*}=IP-\frac{R_{M}}{(n-\delta_l(n))^2},
	\label{Ritz_1}
	\end{equation}
	Here $E_n$ is the level energy, $IP$ is the ionization potential, $R_{M}$ is the mass-reduced Rydberg constant for $^9$Be. The quantum defect $\delta_l(n)$ can be expressed as:
	\begin{equation}
	\delta_l(n)=\delta_{l,0}+\frac{a_l}{(n-\delta_{l,0})^2}+\frac{b_l}{(n-\delta_{l,0})^4}+\cdots
	\label{Ritz_2}
	\end{equation}
	The constant $\delta_{l,0}$, $a_l$ and $b_l$ are all dependent on the orbital angular momentum $l$. At large principal quantum number $n$, all the high order terms vanish. Therefore $\delta_l(n)\approx\delta_{l,0}$ and stays almost constant, as shown in Fig.~\ref{Be Lu}. The NIST data \cite{NIST} at low $n$ region shows core-penetration effects for the $ns$ series and core-polarization effects for the $np$ and $nd$ series. To avoid the complexity of physics modeling at low-$n$, only the high-$n$ levels measured in this work were used into the IP calculation. Therefore the second and higher order terms in Ritz expansion were not required in the fittings. 

	\begin{figure}[!h]
		\begin{center}
			\includegraphics[width=0.7\textwidth]{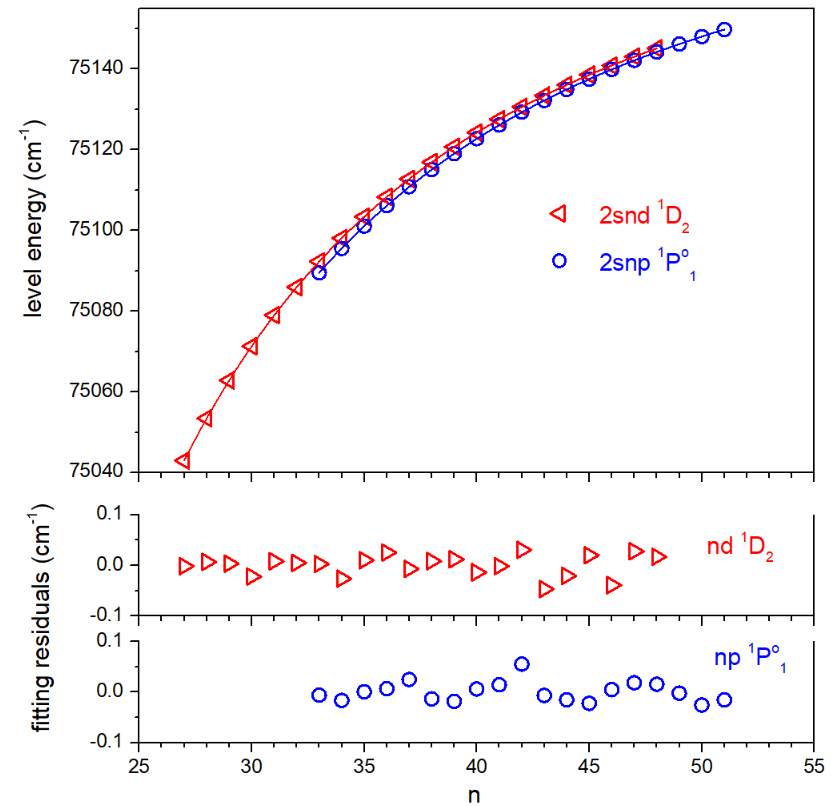}
		\end{center}
		\centering
		\caption{Extracted IP values from two different series $2snd$ $^1D_2$ and $2snp$ $^1P^{\circ}_1$. The fitting residuals were shown in the lower subplots. The weighted average of IP from these two sets of data is 75192.59(3)~cm$^{-1}$ } 
		\label{IP}
	\end{figure}
	
	The IP fits of the series are shown in Fig.~\ref{IP}. The $nd$ series observed from scheme A gave the IP values 75192.55(5)~cm$^{-1}$; and the $np$ series observed from schemes B gave the IP value 75192.61(3)~cm$^{-1}$. The uncertainty includes the systematic error from wavelength measurements, which is 0.02~cm$^{-1}$ for fundamental laser scans and 0.04~cm$^{-1}$ for second harmonic scans \cite{Yuan_Te, LI_Te}. The weighted average of measured IP values is 75192.59(3)~cm$^{-1}$. 
	
	The Be sample used in the off-line experiments is the stable isotope $^9$Be, which has a nuclear spin 3/2. Theoretically the polarization effect on this isotope should be significantly diminished by the hyperfine interaction. As calculated by Eq.(\ref{eq2}), the ratio between parallel and perpendicular polarization ought to be $\sim$2. However the off-line experimental results showed much higher ratios as shown in Fig.~\ref{AI} and \ref{Rydberg}. More investigation is needed to understand this phenomenon.  

	\begin{table*}[!h] \footnotesize
		\caption{$^{9,10,11}$Be RILIS ion yield results (difference between laser on and off) using different laser polarizations at TRIUMF ISAC online facility. The yields of $^{9,10}$Be were measured with a Faraday cup, and those of $^{11}$Be were with a channel electron multiplier. The isotopes were extracted from a UCx target bombarded with 9.8 $\mu$A 480 MeV proton beam. The temperature of the target and ionizer were $\sim$1950~$^{\circ}$C and $\sim$2000~$^{\circ}$C respectively.}
		\begin{center}
			\begin{threeparttable}
				\setlength{\tabcolsep}{7pt}
				\renewcommand{\arraystretch}{1}
				\begin{tabular}{cccccccccc}
					\hline\hline
					& &\multicolumn{8}{c}{isotopes}\\
					\cline{3-10}					
					& laser		& \multicolumn{2}{c}{$^9$Be (I=3/2)}&	&  \multicolumn{2}{c}{$^{10}$Be (I=0)}	& & \multicolumn{2}{c}{$^{11}$Be (I=1/2)}	\\	
					\cline{3-4}\cline{6-7}\cline{9-10}
					& polarization	&	ion yield & ratio\tnote{a} &&ion yield & ratio\tnote{a} &&ion yield & ratio\tnote{a} 	\\
					\hline
					\multirow{2}{*}{theory}	&	 $\lambda_1$ $||$ $\lambda_2$	&	0.22& \multirow{2}{*}{2}&	&	0.11& \multirow{2}{*}{$\infty$}	&&	0.12& \multirow{2}{*}{2.4}	\\
					&	$\lambda_1$ $\perp$ $\lambda_2$	&	0.11&	&&	0&	&&	0.05	\\
					\hline
					\multirow{2}{*}{Nov 30}&	 $\lambda_1$ $||$ $\lambda_2$	&	160$\pm$10 pA& \multirow{2}{*}{5.4$\pm$1.1}	&&	&&	&	62$\pm$10 kcps &\multirow{2}{*}{2.9$\pm$0.7}	\\
					&	$\lambda_1$ $\perp$ $\lambda_2$	&	30$\pm$6 pA	&&		&&&	&22$\pm$4 kcps&	\\
					\hline
					\multirow{2}{*}{Dec 10}&	 $\lambda_1$ $||$ $\lambda_2$	&	208$\pm$15 pA& \multirow{2}{*}{4.5$\pm$0.5}	&&	197$\pm$15 pA	&\multirow{2}{*}{7.9$\pm$1.1}&&953$\pm$ 45 kcps&\multirow{2}{*}{4.4$\pm$0.3}	\\
					&	$\lambda_1$ $\perp$ $\lambda_2$	&	46$\pm$4 pA	&& &	25$\pm$3 pA	&&&	232$\pm$ 15 kcps	\\			
					\hline\hline
				\end{tabular}
				\begin{tablenotes}
					\item[a] the ion yield ratio of $\lambda_1$ $||$ $\lambda_2$ to $\lambda_1$ $\perp$ $\lambda_2$.			
				\end{tablenotes}
			\end{threeparttable}
		\end{center}
		\label{table_online}
	\end{table*}

	\section{Online radioactive Be isotope yield dependence on laser polarization}\label{online}
	We applied the high efficiency Be AI resonance ionization scheme of $2s^2$ $^1$S$_0$$\rightarrow$$2s2p$ $^1$P$^{\circ}_1$$\rightarrow$$2p^2$ $^1$S$_0$ to our on-line RILIS source in Nov/Dec 2018. A uranium carbide (UCx) target was irradiated by a 9.8~$\mu$A proton beam to produce radioactive isotopes of Be. The $\lambda_1$ had 39~mW and $\lambda_2$ had 20~mW, in which the 1st excitation step was well saturated whilst the 2nd step was not. For different Be isotopes, the wavelength of $\lambda_1$ was optimized on the ion signal to compensate for optical isotope shifts. The ion currents or counts were measured downstream by a Faraday cup or a channel electron multiplier respectively, dependent on the beam intensities. The results are shown in Tab.~\ref{table_online}. Considering dark counts/contamination on the detectors and surface ionized isobars in the beam, the ion yields presented here are the difference between laser on and off. For $^9$Be and $^{10}$Be measured by the Faraday cup, hardly any laser-off background was observed. The $^{11}$Be measurements had some laser-off background mainly from radioactive beam contamination on the channel electron multiplier. Surface ionized isobars were barely observed. 
	
	The target temperature was estimated at 1950~$^{\circ}$C from the operating target heating current via the temperature-current calibration and the calculated power deposition from the proton beam. The ionizer was kept at the typical operation current of 230~A, which heated the ionization region up to a temperature around 2000~$^{\circ}$C. The temperatures of the target and ion source are higher than those in the off-line experiment. The laser polarization dependence is still conspicuous although not as significant as in the off-line results. The diminution may be related to an escalating depolarization with the increased hot-cavity temperature. 
	
	The experimental results are compared with the theoretical estimation given by Eq.(\ref{eq2}) for different isotopes with different nuclear spins. The results agree with each other considering the uncertainty of the ion yield measurement, which can be seen from the scattering of the data at different days, and the increased depolarization in the high temperature ion source.

	\section{Summary}\label{summary}
	
	Polarization dependence of resonance laser ionization was investigated at both TRIUMF's off-line laser ion source test stand with stable isotopes and on-line ISAC facility with radioactive isotopes. The results verified the presence of this laser polarization effect in a high atom density environment of hot-cavity sources. Therefore laser polarization is proven to be an important parameter for laser ion source operation. The theoretical calculations gave a fair prediction to the experiments, and should be used as a general guidance when considering the polarization effect on other resonant ionization schemes and elements. Further investigation of elements such as Se and Hg will be pursued. Additionally it is promising to apply this technique to online in-source spectroscopy to help the classification of new structures of rarely studied elements with only radioactive isotopes, such as Ra, Pu and No.  
	
	Besides the intended polarization dependence study, the work also provides a precise measurement of the broad 2p$^2$ $^1$S$_0$ AI state, and the IP derived from Rydberg spectra of $2sns$ $^1S_0$, $2snd$ $^1D_2$ and $2snp$ $^1P^{\circ}_1$ series. Beryllium as the simplest atom with two electrons outside a closed shell is a model system to study electron correlations. The refinement and extension on the electronic structure and spectroscopic knowledge of beryllium atom can be useful for theoretical atomic study and fundamental atom modeling.

    \begin{acknowledgements}
		The experimental work is funded by TRIUMF which receives federal funding via a contribution agreement with the National Research Council of Canada and through a Natural Sciences and Engineering Research Council of Canada (NSERC) Discovery Grant (SAP-IN-2017-00039). M. Mostamand acknowledges funding through the University of Manitoba graduate fellowship.		
	\end{acknowledgements}

	%
	%


\begin{thebibliography}{35}
	
	
	\bibitem{Bekov81} G. I. Bekov, A. N. Zherikhin, V. S. Letokhov, V. I. Mishin, and V. N. Fedoseev, JETP 33 (1981) 450-453. \url{http://www.jetpletters.ac.ru/ps/1509/article_23058.pdf} 
	
	\bibitem{Santala} M. I. K. Santala, A. S. Daavittila, H. M. Lauranto and R. R. E. Salomaa, Appl. Phys. B 64 (1997) 339-347. \doi{10.1007/s003400050182}
	
	\bibitem{Locke} C. R. Locke, T. Kobayashi, T. Nakajima and K. Midorikawa, Appl. Phys. B 122 (2016) 246. \doi{10.1007/s00340-016-6508-7}
	
	\bibitem{Seema} A. U. Seema, A. D. Rath, P. K. Mandal and Vas Dev, Appl. Phys. B 118 (2015) 505–510. \doi{10.1007/s00340-015-6023-2}
	10.1103/PhysRevA.80.052505
	\bibitem{Nit} K. Nittoh, K. Nakayama, J. Watanabe, H. Adachi, H. Ueda and T. Yoshida, J. Phys. B: At. Mol. Opt. Phys. 27 (1994) 1955-1964. \doi{10.1088/0953-4075/27/10/006}
	
	\bibitem{Kim} J. B. Kim, X. Xiong, N. M. Laham, T. B. Lucatorto and T. McIlrath, J. Phys. B: At. Mol. Opt. Phys. 27 (1994) 2953-2960. \doi{10.1088/0953-4075/27/14/028}
	
	\bibitem{Yi} J. Yi, J. Lee and H. J. Kong, Phys. Rev. A 51 (1995) 3053-3057. \doi{10.1103/PhysRevA.51.3053} 
	
	\bibitem{Miyabe} M. Miyabe, M. Oba and I. Wakaida, J. Phys. B: At. Mol. Opt. Phys. 31 (1998) 4559-4571. \doi{10.1088/0953-4075/31/20/014}
	
	\bibitem{Yu89} A. Yu. Elizarov and N. A. Cherepkov, Sov. Phys. JETP 69 (1989) 695-699. \url{http://www.jetp.ac.ru/cgi-bin/dn/e_069_04_0695.pdf} 
	
	\bibitem{Lav13} J. P. Lavoie, R. Li, P. Bricault, J. Lassen, O. Chachkova, and A. Teigelh\"{o}fer, Rev. Sci. Instrum. 84 (2013) 013306. \doi{10.1063/1.4788938} 
	
	\bibitem{Li13} R. Li, J. Lassen, A. Teigelh\"ofer, J. P. Lavoie, P. Bricault, O. Chackakova, J. Meissner and Y. Zlateva, Nucl. Instrum. Methods Phys. Res., B 308 (2013) 74-79. \doi{10.1016/j.nimb.2013.05.005} 
	
	\bibitem{Baig09} M. A. Kalyar, M. Rafiq and M. A. Baig, Phys. Rev. A 80 (2009) 052505. \doi{10.1103/PhysRevA.80.052505} 
	
	\bibitem{Tei10} A. Teigelh\"{o}fer, P. Bricault, O. Chachkova, M. Gillner, J. Lassen, J. P. Lavoie, R. Li, J. Meissner, W. Neu, K. Wendt, Hyperfine Interact. 196 (2010) 161-168. \doi{10.1007/s10751-010-0171-x} 
	
	\bibitem{Li_optics} R. Li, J. Lassen, S. Rothe, A. Teigelh\"{o}fer, and M. Mostamand, Opt. Express 25 (2017) 1123-1130.  \doi{10.1364/OE.25.001123 }
	
	\bibitem{Seb14} S. Raeder, H. Heggen, J. Lassen, F. Ames, D. Bishop, P. Bricault, P. Kunz, A. Mj\o{}s and A. Teigelh\"ofer, Rev. Sci. Instrum. 85 (2014) 033309. \doi{10.1063/1.4868496} 
	
	
	\bibitem{Yakonov_1} M. I. D'yakonov and V. I. Perel', Sov. Phys. JETP 20 (1965) 997-1004. \url{http://www.jetp.ac.ru/cgi-bin/dn/e_020_04_0997.pdf} 
	
	\bibitem{Yakonov_2} M. I. D'yakonov and V. I. Perel', Sov. Phys. JETP 21 (1965), 227-231. \url{http://www.jetp.ac.ru/cgi-bin/dn/e_021_01_0227.pdf} 
	
	\bibitem{Fano65} U. Fano and J. W. Cooper, Phys. Rev. 137 (1965) A1364-A1379. \doi{10.1103/PhysRev.137.A1364} 
	
	\bibitem{NIST} A. Kramida, Yu. Ralchenko, J. Reader and NIST ASD Team. NIST Atomic Spectra Database (ver. 5.6.1), National Institute of Standards and Technology, Gaithersburg, MD, \url{http://physics.nist.gov/asd} (2019).
	
	\bibitem{Clark85} C. W. Clark, J. D. Fassett, T. B. Lucatorto, L. J. Moore and W. W. Smith, J. Opt. Soc. Am. B 2 (1985) 891. \doi{10.1364/JOSAB.2.000891} 
	
	\bibitem{Bochchiat} M. A. Bouchiat and L. Pottier, Journal de Physique Lettres, 36 (1975), 189-192. \doi{10.1051/jphyslet:01975003607-8018900}.
	
	\bibitem{Kuhn} H. G. Kuhn, Atomic spectra (Longmans, Green $\&$ CO. LTD, London and Harlow, 1969) p. 175.
	
	\bibitem{Beigang83} R. Beigang, D. Schmidt and P. J. West, J. Phys. Colloques 44 (1983) C7-229-C7-237. \doi{10.1051/jphyscol:1983719}.
	
	\bibitem{Seaton} M. Seaton, J. Phys. B, 9 (1976) 3001-3007.  \doi{10.1088/0022-3700/7/18/017}. 
	
	\bibitem{Johansson} L. Johansson, Phys. Scr. 10 (1974) 236-240. \doi{10.1088/0031-8949/10/5/008}. 
	
	\bibitem{LiSb17} R. Li, J. Lassen, J. Ruczkowski, A. Teigelh\"ofer and P. Bricault, Spectrochim. Acta B. 128 (2017) 36.  \doi{10.1016/j.sab.2016.12.001}.
	
	\bibitem{LiLu17} R. Li, J. Lassen, Z. P. Zhong, F. D. Jia, M. Mostamand, X. K. Li, B. B. Reich, A. Teigelh\"{o}fer and E. Yan, Phys. Rev. A, 95 (2017) 052501.  \doi{10.1103/PhysRevA.95.052501}.
	
	\bibitem{Seb18} S. Raeder, H. Heggen, A. Teigelhöfer and J. Lassen. Spectrochim. Acta B. 151 (2019) 65-71. \doi{10.1016/j.sab.2018.08.005}.
	
	
	\bibitem{Yuan_Te} T. Kieck, Y. Liu, D. W. Stracener, R. Li, J. Lassen and K. D. A. Wendt, Spectrochim. Acta B. 159 (2019) 105645. \doi{10.1016/j.sab.2019.105645}.
	
	\bibitem{LI_Te} R. Li, Y. Liu, M. Mostamand, T. Kieck, K. D. A. Wendt and J. Lassen, Phys. Rev. A 100 (2019) 052510. \doi{10.1103/PhysRevA.100.052510}. 
	
\end{thebibliography}
	

\end{document}